\documentclass[runningheads]{llncs}
\usepackage[T1]{fontenc}
\usepackage{graphicx}
\usepackage{amsmath}
\usepackage{geometry}
\usepackage{url}
\usepackage{caption}
\usepackage{xcolor}
\usepackage{soul}
\begin{document}
\newgeometry{left=3cm,bottom=3.5cm, right=3cm, top=3.5cm}
\title{A short note on the paper `Are Randomized Caches Really Random?'\cite{chakraborty2023are}}
%
%
\author{Anirban Chakraborty\inst{1} \and Sarani Bhattacharya\inst{2} \and Sayandeep Saha \inst{3}\and Debdeep Mukhopadhyay\inst{1}}
\authorrunning{A. Chakraborty, S. Bhattacharya, S. Saha and D. Mukhopadhyay}
\titlerunning{A short note on the paper `Are Randomized Caches Really Random?'}
%
\institute{Indian Institute of Technology, Kharagpur \\   \email{anirban.chakraborty@iitkgp.ac.in, debdeep@cse.iitkgp.ac.in} \and
Imec, Belgium \email{sarani.bhattacharya@imec.be} \and
Nanyang Technological University, Singapore \email{sayandeep.iitkgp@gmail.com}
}

\maketitle              
\begin{abstract}

In this paper, we analyse the results and claims presented in the paper \emph{`Are Randomized Caches Truly Random? Formal Analysis of Randomized Partitioned Caches'}, presented at HPCA conference 2023. In addition, we also analyse the applicability of `Bucket and Ball' analytical model presented in MIRAGE~\cite{saileshwar2021mirage} for its security estimation. We put forth the fallacies in the original bucket and ball model and discuss its implications. Finally, we demonstrate a cache occupancy attack on MIRAGE with just $10\%$ of total cache capacity and extend the framework to establish a covert channel and a template-based fingerprinting attack.

\end{abstract}
\section{Introduction}
Cache attacks are one of the most potent side-channel attacks on microarchitecture, used to extract sensitive information from a system by observing the pattern in which data is stored and accessed in the cache. A wide variety of attacks can be found in the literature that can leak critical information like encryption keys, user data in the cloud, etc. Although cache attacks have been known to target all levels of the cache hierarchy, the attacks on the last level cache (LLC) bear more significance as they are typically shared across multiple cores; thereby, multiple processes share the LLC at the same time. One of the well-known techniques in cache attacks is the method of evicting a specific cache set with attacker's data to infer security-critical information about the target victim process. Quite naturally, a number of countermeasures have been proposed over the years, which include cache partitioning and randomization of address-to-set mappings. 
Cache randomization and partitioning slows the process of eviction set formation by obfuscating the address-to-set mapping and further dividing the cache logically into multiple partitions. However, prior attacks in literature have shown that newer and more improved algorithms can be envisaged that can discover eviction sets in the secured randomized caches. Another point to note is that although other variants of cache attacks exist which are equally malicious and potent, cache protection schemes, in general, focus more on thwarting eviction-based attacks. 

In~\cite{chakraborty2023are}, published in HPCA'23, we provided a systematic framework to analyse different types of secure cache designs proposed in the literature. The paper first categorized different types of cache designs into four broad classes based on the extent of non-determinism and randomness of allocating an address in those caches. This categorization helps in ``grouping'' of many cache designs based on their randomization feature and design rationale. We then develop a mathematical framework to formally analyse the security implications of the randomized and partitioned cache
designs in terms of collision probability, self-collision probability
and size of the eviction set required to perform a successful
eviction-based attack. Using this mathematical framework, we show the probabilities of \emph{self-collision} between the elements of eviction set of the attacker, that constitutes \emph{noise} in the overall process. In particular, we show that using pre-calculated values for self-collision probability, the number of elements in the eviction set, an adversary can create curated eviction set for all known categories of cache designs.

\section{Regarding Claims on MIRAGE~\cite{saileshwar2021mirage}}
\label{sec:claims}
Along with the aforementioned contributions, the paper~\cite{chakraborty2023are} also  demonstrates the first \emph{set associative eviction} (SAE) phenomenon on a newly proposed randomized design named MIRAGE~\cite{saileshwar2021mirage}. Using the SAE events as signals for eviction, the paper also proposes an algorithm for the generation of eviction set. In the paper, we first show that the Bucket and Ball model proposed in the original paper of MIRAGE~\cite{saileshwar2021mirage} does not take into account all the events that can lead to a \emph{bucket spill} phenomenon. The original paper of MIRAGE assumes a bucket and ball scenario where  each cache set (along with the extra tags provisioned in MIRAGE) represents a bucket and ball represent the number of incoming address lines to be installed in the cache. 

To estimate the probability of bucket spills (a bucket
containing more than $N$ balls), the authors in MIRAGE model the
bucket-state as a Birth-Death chain to calculate the probability
of a bucket containing $N$ balls transitioning into $(N +1)$ balls.
We refer these bucket-states as $N^{th}$ state and $(N +1)^{th}$ state
respectively. Considering the rate of transition from $N^{th}$ state
to $(N+1)^{th}$ state and vice-versa as equal, the authors arrive at
the following equation for bucket spill
\begin{equation}
    \small
    Pr(n=N+1) = \frac{B}{\mathcal{B} \times (N + 1)} \times Pr(n=N)^2
    \label{eqn:p_reduced_n1}
\vspace{-1mm}
\end{equation}
where $\mathcal{B}$ is the number of buckets and $B$ is the total number of
balls thrown.

In~\cite{chakraborty2023are}, we explain that while the event that a particular bucket in $N^{th}$ state transitioning into $(N+1)^{th}$ state occurs with very low probability, it is a restrictive assumption for an attacker. As an attacker, it is free to fill up any one of the buckets, thereby biasing the cache
towards a more vulnerable state. Therefore, using Coupon Collector's problem, the probability that any
bin containing exactly $N + 1$ balls can be given by
\begin{equation}
    \small{
\begin{aligned}
    Pr[n = N+1] &= \binom{B}{N+1} \left( \frac{1}{\mathcal{B}}\right) ^{N+1} \left (1 - \frac{1}{\mathcal{B}} \right) ^{B - (N+1)} \times \mathcal{B} \\ &
    \approx \frac {\lambda ^ {N+1} e^{-\lambda}}{(N+1)!} \times \mathcal{B} \qquad where \quad \lambda = \frac{\mathcal{B}}{B}
\end{aligned} }%
\label{eqn:pr_n1_j_bin}
\end{equation}

\begin{figure}[!t]
        \centering
        \includegraphics[scale=0.5]{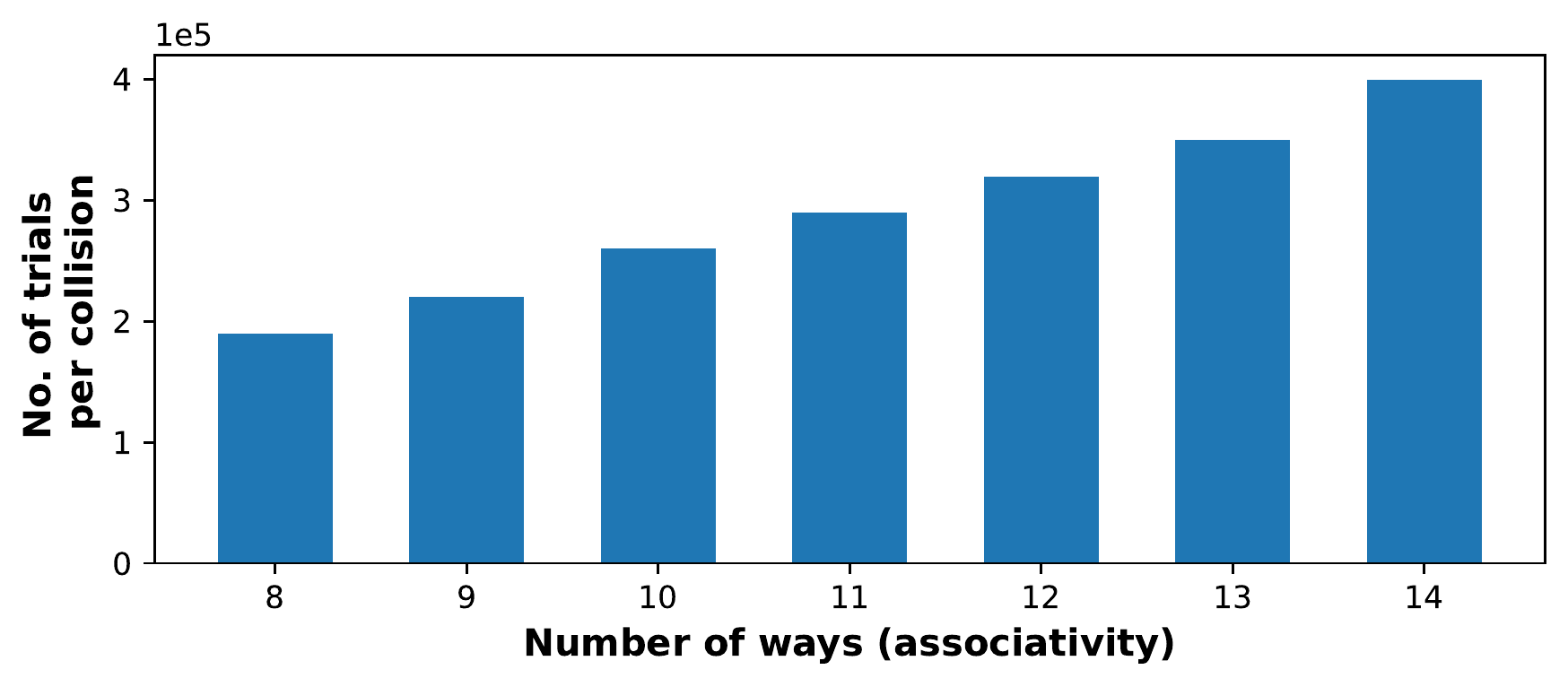}
        \caption{Frequency of collisions in MIRAGE, with associativity varying from $8$ to $8+6$~\cite{chakraborty2023are}}
        \label{fig:bucket_spill_mirage}
\end{figure}

\begin{figure}[!t]
    \centering
    \includegraphics[scale=0.6]{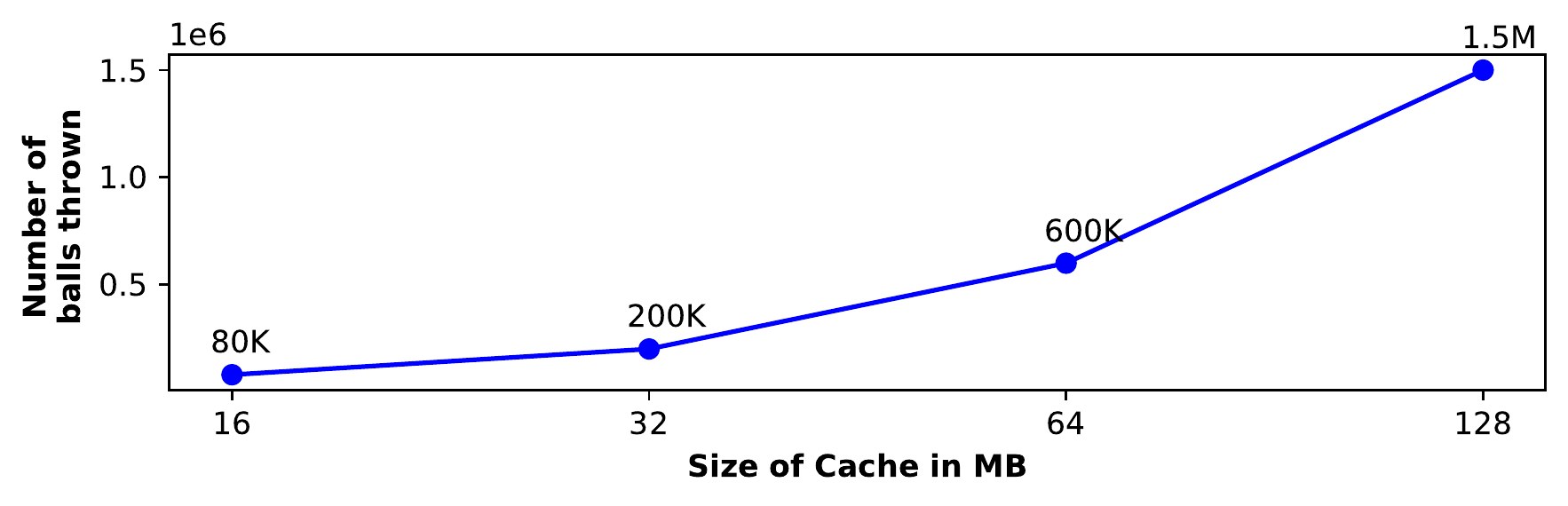}
    \caption{Number of balls thrown for atleast one bucket spill for different size of the cache~\cite{chakraborty2023are}}
    \label{fig:bucket_spill_trend}
\end{figure}

In the above analysis, the attacker starts from an empty cache and randomly tries to enforce a bucket spill in any of the $\mathcal{B}$ buckets. However, we also note in~\cite{chakraborty2023are} that even if the initial state of the cache is not
considered as completely empty, it is still possible to deplete
all the ways of a sibling set (a bucket spill). Using Eqn.~\ref{eqn:pr_n1_j_bin}, we showed empirical results on an in-house MIRAGE LLC simulator that bucket spills are observable in around $10^5$ - $10^6$ ball throws with the cache parameters set as the recommended version on MIRAGE (Fig.~\ref{fig:bucket_spill_mirage},~\ref{fig:bucket_spill_trend}). 



\section{Technical Problems in the Claim}
While the mathematical model discussed in Section.~\ref{sec:claims} holds true for the Bucket and Balls Analysis, we found a critical problem arising due to incorrect representation of MIRAGE. As per the original paper of MIRAGE~\cite{saileshwar2021mirage}, the authors claim that the bucket and ball model properly represents the working principle of MIRAGE. Their claim on the security guarantees of MIRAGE rests upon the validity of the Bucket and Ball Model. In this article, we point out a major flaw in the representation of the original Bucket and Ball model and explain why such a model does not properly justify the abstractions of a practical cache design. But before we move forward, we provide a brief background on the design of MIRAGE for brevity. 

\subsection{Background on Mirage}
MIRAGE~\cite{saileshwar2021mirage} is a randomized cache design proposed in USENIX Security 2021. MIRAGE's design rationale differs from other categories of randomized and partitioned cache architectures. Traditional as well as the majority of well-known randomized cache designs, map any
particular address only to a small set of cachelines, as the randomized addresses are still mapped as per set-associative pattern. Even
with the introduction of partitioning, the total number of
cachelines that an address can be mapped to remains fixed. MIRAGE tries to introduce a \emph{``pseudo''} \emph{fully associative} randomized cache design thereby making it difficult to construct an eviction-set. MIRAGE decouples the tag-store and data-store (as shown in Fig. 3) and uses set-associative
lookup in tag-store while maintaining fully associative nature
of the data store. 

\begin{figure}[!h]
    \centering
    \includegraphics[scale=0.6]{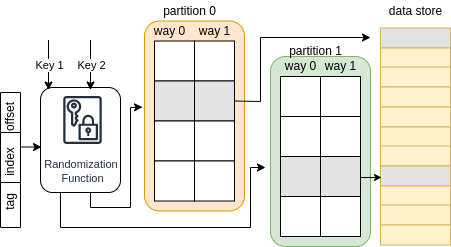}
    \caption{MIRAGE~\cite{saileshwar2021mirage}. Set associative Tag store and Fully Associative Data Store}
    \label{fig:mirage_design}
\end{figure}

The central idea is to avoid set-conflicts
by enforcing global replacement on a cache miss. To enable
global evictions, MIRAGE provisions extra invalid tags in each
set in the tag store. Thus, on arrival of a new address for
mapping into the cache, an invalid tag from the tag-store
can be used to map the address and simultaneously an entry
from the data store is chosen randomly, thereby providing
global eviction. In addition to the global eviction provisioning
and decoupling of tag and data store via pointer indirection,
MIRAGE also uses the concept of skewed caches along with
randomization of the input address. The tag-store is split into
two skews while having a single fully associative data-store.

\setcounter{footnote}{0} 
\subsection{Security Analysis through Bucket and Ball Model}
To provide a security analysis of their scheme, the authors in MIRAGE paper~\cite{saileshwar2021mirage} used an abstraction by modelling each set of tag-store as a bucket and each address installation as a ball throw. However, the analytical model they proposed (as shown in Eqn.~\ref{eqn:p_reduced_n1}) does not properly represent the working principle of the design. The authors model the behaviour of bucket and ball system as a Birth-Death chain~\cite{saileshwar2021mirage}. Therefore, the analysis in Mirage essentially presents a scenario where a bucket in its $N^{th}$ state moves to $(N+1)^{th}$ state. In~\cite{chakraborty2023are}, we show that such probability is not significantly low, as claimed in MIRAGE, since in this case, the attacker can target to find \emph{any} bucket transitioning from $N^{th}$ to $(N+1)^{th}$ state. 
\textcolor{black}{\textit{However, this modelling of MIRAGE is inappropriate.} One might recall that MIRAGE provisions extra invalid tags in each set. Furthermore, it uses the concept of load balancing where any incoming address can be mapped to two \emph{sibling} sets. Therefore, considering $m$ extra ways in each skew, the attacker needs to fill all $m$ ways in both the sets in order to achieve a collision.
}

Let us understand this scenario by taking a reference MIRAGE design. Consider $16$MB MIRAGE cache with $8+6$ ways in the tag store~\footnote{The default version of Mirage requires $8$ base and $6$ extra tags in each set.}. Therefore, the tag store is indexed in a set-associative way with $16K$ sets in each skew (total $16K$ sets). Likewise, the number of entries in the data-store is $16K \times 8$, assuming $8$ valid ways in each set. Keeping in line with the assumptions of Bucket and Ball model from the original MIRAGE paper~\cite{saileshwar2021mirage}, we initialize the cache with some valid and some invalid tags. Given an initial state with $x$ valid and $y$ invalid tags in
each set, the attacker accesses arbitrary addresses to map them into the sibling sets. For a $64$-bit addressing system, the input address is first encrypted using two fixed keys producing two independent $64$-bit ciphertexts $\mathcal{C}_1$ and $\mathcal{C}_2$. According to the load-balancing principle of MIRAGE, the index bits of both $\mathcal{C}_1$ and $\mathcal{C}_2$ are used to index into the two  \emph{sibling sets}. As there are $16K$ sets, $14$ bits from the ciphertexts are used to map into corresponding sets in each skew. In other words, an incoming address creates a unique tuple ($i_1, i_2$) after encryption where $i_1$ indicates the set number in skew $1$ and $i_2$ indicates the set number in skew $2$.  Therefore, in order to get a collision, an adversary must choose an address such that its corresponding ciphertexts have $14$ bits of ($i_1, i_2$), total $28$ bits, matches with the elements that are already present in those sibling sets. \textcolor{black}{We now show that probability of finding such a ``collision'' is not negligible (contrary to the claims made in MIRAGE~\cite{saileshwar2021mirage}).}


\subsection{Collision using Birthday Paradox}~\label{sec:birthday}
Birthday Paradox~\cite{suzuki2006birthday} is a statistical phenomenon that shows the probability that any two events having a collision increases to nearly $50\%$ if the number of samples is significantly large. In case of MIRAGE, one needs to match $28$ bits from a random distribution (assuming uniform distribution of ciphertext by the encryption function) to find a collision in a selected tag set. However, as per the Birthday Paradox, the adversary can find a pair of indices ($j_1, j_2$) that collide in \emph{any} of the sets in $2^{\frac{28}{2}} = 2^{14}$ accesses. Applying this relationship to the bucket and ball problem, the adversary needs $2^{14}$ accesses to transition any bucket residing in  $N^{th}$ state into $(N + 1)^{th}$ state. 
However, MIRAGE provisions extra invalid tags in each tag store. For our example cache, each set will have $6$ extra invalid tags, thereby total number of invalid tags in a pair of sibling sets will be $12$. Therefore, to enforce a set associative eviction, the adversary needs a $12$-way collision in one of the sibling sets. \textcolor{black}{While the analytical model for bucket and ball models the event when a bucket transitions from $N^{th}$ to $(N+1)^{th}$ state, the correct model should consider the event where any bucket from $N^{th}$ state transitions into $(N + m)^{th}$ state, $m$ being the number of extra tags in each tag set. We conclude that the analytical results presented in both ~\cite{saileshwar2021mirage} and \cite{chakraborty2023are} do not properly represent the cacheline assignment operations of MIRAGE. Both the papers attempts to model the ``bucket spill'' event as transitioning from  $N^{th}$ to $(N+1)^{th}$ state, while the correct transition should be from $N^{th}$ to $(N+m)^{th}$ state. Such a transition requires $m$-way collision (not a single collision) that is statistically improbable to achieve, thereby providing it adequate protection against \emph{collision-based} attacks. 
}

\section{Regarding the results on MIRAGE in \cite{chakraborty2023are}}
In the previous section, we discussed about the fallacies in the bucket and ball model of MIRAGE, both in papers \cite{saileshwar2021mirage} and \cite{chakraborty2023are}. However, the results on MIRAGE in \cite{chakraborty2023are}, specifically Fig. 7 and Fig. 10 (from paper~\cite{chakraborty2023are}), were produced by an in-house simulator of MIRAGE and not bucket and ball model abstraction. Both these plots signify that a set associative eviction is possible in Mirage, which can eventually be used to create an eviction set. Unfortunately, we found out that the results were faulty due to an incorrect implementation of the randomizing block cipher. The in-house simulator uses PRESENT block cipher~\cite{bogdanov2007present} for randomizing the input addresses using two distinct keys. PRESENT in a lightweight block cipher which has been proven to cryptanalytically robust against all known attacks. The design of MIRAGE assumes that the cipher chosen must ensure that the addresses are uniformly mapped to cache sets. For the in-house simulator, we used a GitHub implementation of PRESENT, written in Python2~\cite{github_present}. Since the rest of the simulator was implemented in Python3, the cipher implementation had to be ported as well. The ported version of the cipher takes as input a binary representation of the address and converts it into an integer for further operations of the cipher. It is during this conversion that the binary representation is inadvertently converted to integer assuming it to be in hexadecimal format. The consequence is that a $64$-bit input gets transformed into $128$-bit while the rest of the cipher operation is done on the last $64$-bits of the converted value. This leads to a situation where the two distinct input addresses $a_1$ and $a_2$ have identical last $16$ bits, which leads to same ciphertext and thereby mapping into the same sibling sets. When resolved, the cipher produced uniform distribution over the set indices and \emph{no collision were observed for a 16MB MIRAGE with $8+6$ ways} in each tag set. This observation also conforms with our argument in  Section.~\ref{sec:birthday} that such an event requires $12$-way collision which has a very low probability.

However, we also note that the paper \cite{chakraborty2023are} proposes a framework to generate efficient eviction sets for a wide variety of secure cache designs. While the results on MIRAGE are incorrect due to non-uniform distribution of the ciphertext, the other results in the paper are not affected by this problem. In \cite{chakraborty2023are}, we develop a mathematical framework for randomized and partitioned cache designs and generate formulations for self-eviction probability and the minimal number of elements required in the eviction set. We use these parameters to develop a generic algorithm that works on all variants of cache designs, except MIRAGE. MIRAGE being a pseudo fully-associative cache, behaves differently from the rest of the available cache designs in the literature. 

\section{Is MIRAGE really secure?}
For a set-associative eviction in MIRAGE, which is key to create an eviction set, an adversary needs to generate $m$-way collisions in any of the tag sets. Therefore, eviction based attacks on Mirage is seemingly impractical due to its provisoning of extra invalid tags. However, eviction based attacks are one of the many cache attacks variants available in literature. It is important to analyse any design from the context of all known and practical attacks to consider a design resonable secure and \emph{non-leaky}. 

\subsection{Global Eviction in MIRAGE}
A major feature of MIRAGE is that it implements the data-store as a fully associative cache. However, there is a significant difference. In a fully associative cache, any incoming address can be placed into any slot, \emph{provided that the slot contains an invalid entry or it is free}. Therefore, for a fully associative cache, the CPU searches through the entire cache for slots containing invalid entries for installing the new incoming address. In case when no such invalid entry is found, one of the valid entries is evicted based on the replacement policy in place. However, fully associative caches are impractical for LLCs as they bear significant performance overhead due to exhaustive search requirement. In order to alleviate that problem, MIRAGE uses a random eviction policy for all entries, which is termed as \emph{global eviction}. For any incoming address, a random entry from the data store is evicted and its corresponding entry from the tag store is invalidated. One must note that the global eviction in MIRAGE is a deterministic event. In other words, a global eviction happens for every new address installation, irrespective of whether the data store contains any invalid entry. This is in stark contrast to both fully associative and other randomized designs where eviction happens only when certain  conditions are met. In this section, we discuss how such seemingly benign event can leak information regarding victim process and in fact introduce newer attack vectors.

\subsection{Cache Occupancy Attack}
Cache attacks can be categorized into three broad classes: access, timing and trace attacks. The eviction based attacks such as Prime+Probe, Evict+Time, etc. tries to fill up certain sets in the cache in order to create collision with the victim process. An important criterion for these types of attacks is that the attacker must fill all the ways in the target cache set in order to leak information about the victim. However, a separate class of cache attacks have surfaced recently that does not depend on the granularity of cache set information. The attacks, broadly called as cache occupancy attacks, focuses on the cache footprint of the victim in terms of cache occupancy, rather than relying on activity within specific cache sets. Such attacks are powerful as they have been used in literature to perform website fingerprinting~\cite{shusterman2019robust,shusterman2020website}. MIRAGE, due to its policy of compulsory eviction from data store for every new address installation, is particularly vulnerable to such cache occupancy attacks. In the next section, we discuss how cache occupancy attacks can lead to covert channel and templating-based fingerprinting in MIRAGE. 

\section{Covert Channel in Mirage}
Covert channel is a type of illicit communication channel that is designed to be hidden or concealed from system detection mechanisms. Covert channels are serious security concerns as they can be used to transmit sensitive information across process boundaries. Since the LLC is shared across all processors, it provides a suitable surface for establishing covert channel across processes running on separate physical cores. In this section, we show a low-error rate covert channel on MIRAGE which can be set up by occupying less then $10\%$ of the entire data store. 

\subsection{Setting up the channel}
To demonstrate the covert channel, we use an in-house LLC simulator for MIRAGE with $16$MB size and $8+6$ ways in each tag set with $2$ skews. Each incoming address is randomized using PRESENT block cipher with two fixed and distinct keys. We note that the PRESENT block cipher implementation passed the test vectors presented in the original paper~\cite{bogdanov2007present}.  Therefore, the number of locations in the data cache is $131,072$ as there are $16$K sets and $8$ valid ways. 

Covert channels are typically unidirectional channels where two parties - a sender and a receiver - collude to transfer information by using a shared resource and pre-decided pattern. To make a covert channel practical and stealthy in order to avoid raising suspicion, the two parties shall not have any explicit communication channel, such as shared memory. Cache memory has been used as a high-bandwidth covert channel in literature before. However, the process requires the sender and the receiver to occupy a major portion of the cache to meaningfully transmit information. In this section, we demonstrate a cache covert channel using cache occupancy attack on MIRAGE by occupying less than $10\%$ of the entire cache memory. Moreover, as every incoming address is mapped randomly (due to randomization by a block cipher), the memory footprint of both parties essentially creates a random access pattern in the cache, thereby raising no suspicion of a covert operation. To elaborate on the working principle of the attack, we discuss the operations undertaken by the two parties below.

\subsubsection{Receiver: } The sender allocates a memory space and randomly accesses $10,000$ addresses. In order to ensure that each address occupies an entire cacheline, the addresses are accessed in a sequential manner with an interval of $1,000$ between each address. For example, if the first address accessed is $p$, then the next address to be accessed will be $1000 + p$. This ensures that none of the addresses experiences a cache hit; thus, each address will occupy an entire cache line. It is important to note that due to the presence of global eviction for each incoming address (irrespective of the occupancy level of the cache), multiple addresses from the set of $10,000$ addresses will be randomly evicted. In our experiments, we discovered that out of $10,000$, around $400$ addresses get evicted at the end of the entire operation. One must note that all the $10,000$ addresses are mapped to different cachelines, and the starting cache state is completely empty. However, the global replacement policy of MIRAGE evicts around $400$ already installed addresses due to random selection from the data cache.
Therefore, after the sender finished accessing its addresses, around $9600$ entries in the data store are filled with valid entries, which is $7.32\%$ of the entire data store capacity.

\subsubsection{Sender: } The receiver is assumed to be executing on a separate physical core, completely agnostic of the addresses accessed by the sender. Moreover, the randomization scheme of MIRAGE anyway scrambles any dependency between two addresses. Once the sender has inserted its own addresses, the receiver starts its operation. Similar to the receiver, the sender allocates a memory space and, depending on the value of the bit ($0$ or $1$) it wants to transmit, either makes $1000$ or $4000$ memory accesses. Additionally, the addresses accessed are at an interval of $1000$ in order to ensure no two accesses are mapped to the same cache line. Due to each insertion of the addresses by the sender, one of the data store entries is evicted as per the global eviction policy. Interestingly, we observed that due to random selection of eviction candidates from the data store, some of the already installed entries of the sender are evicted as well. To put in perspective, when the sender makes 1000 accesses, an average of $90$ addresses are evicted at the end of the operation. While for $4000$ accesses, an average of $350$ addresses of the sender are evicted.

\subsubsection{Receiver: } Once the sender has completed its operation, the receiver now re-accesses all the $10,000$ addresses it has installed in the cache. It is worth mentioning that the receiver at this point does not have any knowledge about the number of its own addresses being evicted from the cache, neither it has knowledge about the number of addresses from the sender installed presently in the cache.
The receiver re-accesses the addresses and measures the access times for each of them individually. The ones that are still in the cache will experience a cache hit, while the ones that have been evicted due to random global evictions will experience a cache miss. In our experiment, we observed that when the sender accesses $1000$ addresses, the receiver observes cache misses for around $450$ of its own pre-installed addresses. Likewise, for $4000$ accesses of the sender, the receiver observes around $750$ cache misses. Therefore, by measuring the total number of cache misses, the receiver is able to determine the secret bit transmitted by the sender.

\section{Template based Fingerprinting}

\begin{figure}[!t]
    \centering
    \begin{minipage}{0.49\textwidth}
        \centering
        \includegraphics[scale=0.55]{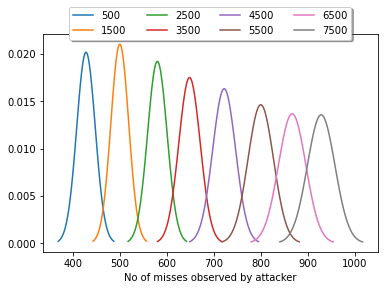}
        \captionof{figure}{Templates of victim accesses with total number of accesses ranging from 500 to 7500 and interval of 1000 accesses}
        \label{fig:500_template}
    \end{minipage}
        \hfill
    \begin{minipage}{0.49\textwidth}
        \centering
        \includegraphics[scale=0.55]{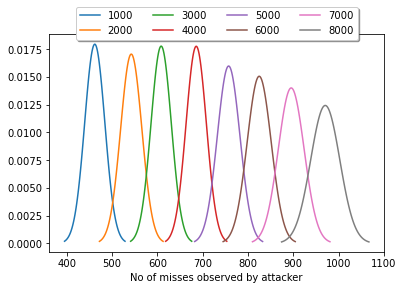}
        \captionof{figure}{Templates of victim accesses  with total number of accesses ranging  from 1000 to 8000 and interval of 1000 accesses}
        \label{fig:1000_template}
    \end{minipage}
\end{figure}

The random selection of data cache entries for global eviction in MIRAGE creates an opportunity for a stealthy covert channel across cores by only occupying less than $10\%$ of the cache. In this section, we extend the attack to perform a more potent side-channel attack by using the cache occupancy exploit. In particular, we show a potential template-based fingerprinting of unknown processes in a system, based on the number of accesses made by them. Website fingerprinting using cache occupancy attacks has been shown in~\cite{shusterman2019robust,shusterman2020website}. Cache occupancy attacks are considered more powerful than contention-based attacks as they have lesser requirements and does not need granular information like other cache attack variants. However, cache occupancy attacks on classical caches as well as most randomized schemes require the attacker to occupy a substantial part of the cache in order to enforce the victim to evict a large number of attacker's addresses. The global eviction of Mirage opens up a new vulnerability which can be exploited to create templates based on the aggregate number of cache misses to perform fingerprinting on processes. Keeping in line with the covert channel experimental setup, the attacker allocates a memory space and accesses $10,000$ addresses to install them in the cache in distinct cachelines. The objective of the attacker is to create templates on varying number of accesses of a hypothetical/dummy victim process. The dummy victim process executes different numbers of cache accesses starting from 1000 till 8000, with an interval of 500. Therefore, we create 16 templates of total accesses, each separated by 500 accesses. Fig.~\ref{fig:500_template} and~\ref{fig:1000_template} shows the templates created for different number of total accesses where the $x$-axis denotes the total number of misses observed by the attacker after the victim has finished its accesses. The lineplots actually denote the distribution of total number of misses observed by the attacker for different accesses of the victim. \textcolor{black}{As evident from the plots, if the cache footprint of two (or more) processes differ by at least 1000 distinct memory accesses, they can be clearly identified based on the templates. This leads to template-based fingerprinting of processes just by measuring the number of cache misses of the adversary's own memory addresses.}

\subsection{Impact of the choice of cipher}
MIRAGE assumes that the set-index derivation function produces output that are uniformly distributed. The templates shown in Fig.~\ref{fig:500_template} and~\ref{fig:1000_template} were created on a simulator having PRESENT as the randomising function with two fixed and distinct keys. In order to ascertain that the choice of key and overall cipher does not have an impact on the cache occupancy attack, we performed the template formation experiment using a different block cipher. Since the authors of Mirage used PRINCE~\cite{borghoff2012prince} block cipher in their analysis, we performed the same template creation experiment using PRINCE cipher with keys randomly selected for each template formation. This shows that the template formation does not depend on the choice of cipher or the key. This observation is quite intuitive as the global eviction is not based on the cipher or the key, rather it selects victim entries from the data store randomly. As long as the randomizing function produces uniformly distributed output, a template created using any arbitrarily chosen key can be used for fingerprinting processes whose addresses are encrypted with a totally different key. This result also signifies that the attack cannot be stopped by rekeying. Once a template is created, the attacker can use it, irrespective of the rekeying period.

\subsection{Cache occupancy on other randomized variants}
An important question at this juncture is whether MIRAGE is comparatively more vulnerable to cache occupancy attacks when compared to other cache variants, including the classical set-associative cache. \emph{Our experiments have answered this question in positive}. In these cache designs, the attacker needs to occupy a large portion of the cache in order to create cache occupancy attack. To observe cache misses, an attacker needs to fillup multiple sets whereas in case of MIRAGE, the global eviction helps in eviction of the attacker addresses.

\section{Future Direction and Conclusion}
The attacks shown in this paper is a proof-of-concept work conducted on an in-house LLC simulator built on Python3, based on the design principles of MIRAGE. The future course of work will be to verify these observations on a standard architectural simulator (like gem5~\cite{binkert2011gem5}) and perform fingerprinting of different processes and workloads.

Cache occupancy attack is a powerful and potentially dangerous attack technique as it does not have strict requirements and can leak information about victim's operations. Majority of the secure cache designs focuses on developing resilience against contention-based attacks while ignoring other attacks. The attack on MIRAGE serves as a stark reminder that designing countermeasures against one class of attack can potentially make the design vulnerable to other attacks.

\section*{Acknowledgement}
We would like to thank the authors of Mirage~\cite{saileshwar2021mirage} for identifying the non-uniform distribution of the PRESENT cipher output, which led us to find the issue with the implementation and the incorrect representation of Bucket and Ball model.

\bibliographystyle{splncs04}
\bibliography{mybibliography}

\end{document}